\documentclass[useAMS,usenatbib]{mn2e}

\usepackage{epsfig}

\def\lesssim{\mathrel{\hbox{\rlap{\hbox{\lower4pt\hbox{$\sim$}}}\hbox{$<$}}}}
\def\gtrsim{\mathrel{\hbox{\rlap{\hbox{\lower4pt\hbox{$\sim$}}}\hbox{$>$}}}}
\newcounter{subfig}

\title[Particle disks near planets]
{Diffusive low optical depth particle disks truncated by planets}

\author[Quillen]
{Alice C. Quillen  \\
Department of Physics and Astronomy, University of Rochester,
Rochester, NY 14627;
aquillen@pas.rochester.edu}

\begin{document}
\label{firstpage}
\maketitle
\begin{abstract}
Two dimensional particle disks in proximity to a planet are numerically
integrated to determine when
a planet in a circular orbit can truncate a particle disk. 
Collisions are treated by giving each
particle a series of velocity perturbations during the integration.
We estimate the mass of a planet required to truncate
a particle disk as a function of collision rate, related
to the disk optical depth, and velocity perturbation size,
related to the disk velocity dispersion.
We find that for particle disks in the regime estimated for debris disks, 
a Neptune mass planet is sufficiently massive to truncate the
disk. If both the velocity dispersion 
and the disk optical depth are low 
(dispersion less than approximately 0.02 in units of circular motion, and 
optical depth less than $10^{-4}$)
then an Earth mass planet suffices. 
We find that the disk is smooth and 
axisymmetric unless the velocity perturbation
is small and the planet mass is of order or greater than
a Neptune mass in which case azimuthal structure is seen
near prominent mean motion resonances.
\end{abstract}


\section{Introduction}

Recent observations made with the Spitzer Space Telescope have
added to the total number of known debris and circumstellar disks
\citep{rieke05,beichman05,gorlova06,chen06,bryden06,beichman06}.
Theses disks have total infrared opacity, as estimated from the fraction
of stellar light re-emitted in the infrared, in the range
$\tau_{IR} \sim 10^{-3} - 10^{-5}$.   
Infrared spectra of many debris
are well fit with a single black body temperature
suggesting that they possess inner holes \citep{chen06}.
Planets are suspected to be responsible for the inner clearings
(e.g., \citealt{kalas05,quillen06,chen06}). 

The fraction of starlight absorbed by the dust is
related to the timescale between dust particle collisions.
Here we focus on collisions as they are expected to
make significant modifications to the dynamics.
The relation between the fraction of starlight absorbed by the dust
and the collision timescale  depends  on the width
and radius of the disk.
A ring with radius $r$, width $dr$, and optical depth normal
to the disk plane, $\tau_n$,
has total surface area $A= 2 \pi r dr \tau_n$ however at radius
$r$ the starlight passes through a sphere of area $4 \pi r^2$.
For low opacity systems
the infrared opacity is related to the disk optical depth  by
\begin{equation}
\tau_n \approx {2r \over dr} \tau_{IR}.
\end{equation}
We expect that the disk optical depth, $\tau_n$, exceeds $\tau_{IR}$ by a modest
factor ($dr<r$) and that $\tau_n$ is the same
order of magnitude as $\tau_{IR}$ unless the ring is 
very narrow.
The infrared opacity measurements allow us to
crudely estimate the normal disk
optical depth and so collision timescale in the disk, though
when possible,
direct imaging in the infrared can provide better measurements
(e.g., \citealt{marsh05}).

Planets have been proposed to explain the morphology of a
 number of systems  with dusty disks
(e.g., \citealt{ozernoy00,wilner02,quillen02,wyatt03,deller05}),
however most models do not take into account collisions between
particles in the disk.   The collision timescale, $t_{col}$,
for the smaller particles can be estimated
from the optical depth, $\tau_n$,  normal to the disk plane, 
\begin{equation}
t_{col} \sim (3 \tau_n n)^{-1}
\end{equation}
\citep{hanninen92}
where $n$ is the mean motion at radius $r$.
As emphasized by \citet{dominik03,wyatt05},
for most debris disks the collision timescale between particles 
is shorter than the Poynting-Robertson drag timescale and so collisions
should be considered when interpreting the evolution and morphology
of these disks.  \citet{quillen06} proposed that the steep inner edge
of Fomalhaut's disk might be due to the removal 
(by the planet) of particles 
scattered interior to the disk edge by collisions.
While \citet{quillen06} suggested that the chaotic zone
near the corotation region is a likely location
for disk truncation,
this has yet to be tested by numerical simulations that incorporate
both perturbations by a planet and
collisions on a timescale consistent with that expected for debris disks.
That is what we aim to do here.

For disk optical depths in the regime of debris
disks, $\tau_n \la 10^{-2}$, spiral density waves 
cannot be driven  at Lindblad resonances 
(\citealt{quillen06}; this regime
is also discussed in the context of planetary
rings by \citealt{franklin80,hanninen92,lissauer98,espresate01}).
However collision induced angular momentum transfer causes a
thin disk to diffuse radially and become wider.
A particle ring has two characteristic timescales \citep{brahic77}, 
the collision timescale, $t_{col}$, that is related to
the disk optical depth, and the diffusion timescale, $t_{diff}$, that
depends on both the disk velocity dispersion and collision timescale.

The radial evolution of an isolated particle ring can be described with 
a diffusion equation 
\begin{equation}
{\partial N \over \partial t}  = 
{\partial  \over \partial r} \left(D {\partial N \over \partial r} \right)
\label{eqn:diffusion}
\end{equation}
where $N(r)$ is the number density as a function of radius $r$,
(e.g., \citealt{petit87} equation 48 and 
\citealt{lithwick06} equation 11). 
The diffusion coefficient, $D$,  depends
on the collision timescale and velocity dispersion, $u^2$, in the disk,
\begin{equation}
D \sim {u^2 \over t_{col} n^2}.
\label{eqn:dcoeff}
\end{equation}
This diffusion coefficient is similar to a viscosity and
can be estimated by considering the mean free path and particle
velocity differences set by the epicyclic amplitude (e.g., \citealt{M+D}).
As the collision time depends on the number density,
the diffusion coefficient can depend on $N$ \citep{petit87,lithwick06}.

\renewcommand{\thefigure}{\arabic{figure}\alph{subfig}}
\setcounter{subfig}{1}
\begin{figure}
\includegraphics[angle=0,width=3.5in]{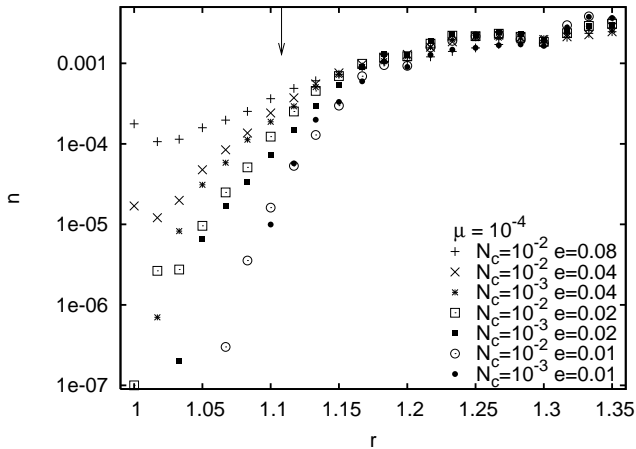}
\caption{
Azimuthally averaged density as a function of radius 
for simulations C1,D1,E1,F1,M1,N1 and L1 listed in table \ref{tab:tab1}.
These simulations have different numbers of collisions
per orbit and velocity perturbations, but the same
planet mass ratio, $\mu = 10^{-4}$.
Radii are given in units of the planet's semi-major axis.
Arrows are drawn at the radii corresponding to the location
of the chaos zone boundary or at $r = 1 + 1.5\mu^{2/7}$ 
for each planet mass ratio that
has a displayed simulation.
\label{fig:rplotp1}
}
\end{figure}

\addtocounter{figure}{-1}
\addtocounter{subfig}{1}
\begin{figure}
\includegraphics[angle=0,width=3.5in]{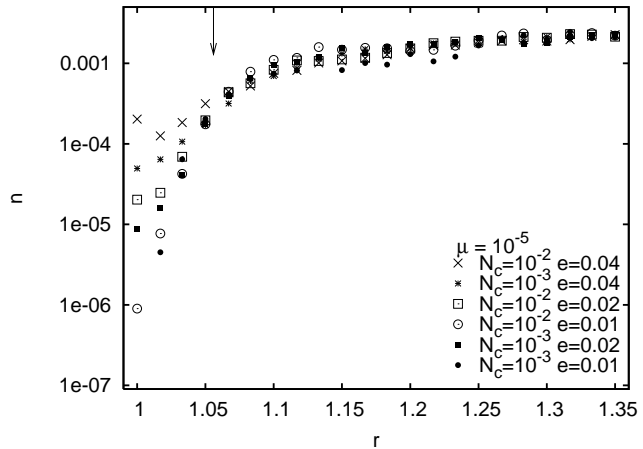}
\caption{
Similar to \ref{fig:rplotp1} except simulations
C2,D2,E2,F2,M2, and N2 are shown with a planet mass ratio of $\mu=10^{-5}$.
\label{fig:rplotp2}
}
\end{figure}

\addtocounter{figure}{-1}
\addtocounter{subfig}{1}
\begin{figure}
\includegraphics[angle=0,width=3.5in]{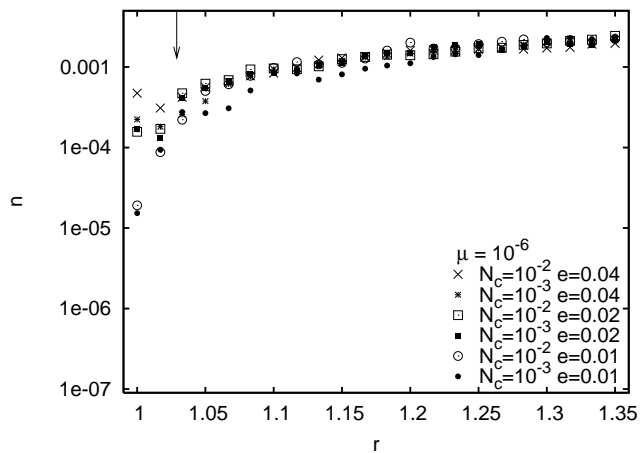}
\caption{
Similar to \ref{fig:rplotp1} except simulations
C3,D3,E3,F3,M3, and N3 are shown with a planet mass ratio of $\mu=10^{-6}$.
\label{fig:rplotp3}
}
\end{figure}

\renewcommand{\thefigure}{\arabic{figure}}

In this paper we adopt a diffusive approach toward simulating
the affect of collisions on a particle disk that resides
near a planet.  
When the disk optical depth is low, many orbits pass
between collisions.  During this time particles can be integrated
directly under the influence of only gravity.  Previous 
works simulating collisions adopt a statistical method 
for predicting when each particle suffers a collision and its affect 
(e.g.,  \citealt{espresate01,melita98}),  or
search for close approaches between particles 
and then compute momentum changes for both particles 
(e.g. \citealt{lewis00,charnoz01,lithwick06}).  
In this work we adopt the first approach. 
We only perturb particles on a mean collision timescale and
compute a perturbation to the particle's momentum.
Both the collision
timescale and momentum perturbation
are chosen independent of the particle distribution (similar
to the approach by \citealt{espresate01}).
A more sophisticated simulation would allow the particle
distribution to set both quantities (e.g., \citealt{melita98}). 
However this requires additional computational
effort, either by computing 
close approaches carefully \citep{charnoz01},
letting the computed collisions depend on a numerical grid 
\citep{lithwick06} or integrating sufficient numbers
of particles that the 
velocity distribution is well sampled \citep{melita98}.  

\section{Numerical integrations}

Numerical integrations were carried out in the plane,
using massless particles under the gravitational
influence of only the star and a
massive planet with zero eccentricity, $e_p=0$, and
semi-major axis, $a_p = 1$, using a
conventional Burlisch-Stoer numerical scheme.
Initial particle 
mean anomalies and longitudes of perihelia were randomly chosen.
In each simulation 1000 particles were integrated for 
$\sim 10^5$ planetary orbital periods.

Each particle suffers collisions with a mean collision rate
set by the collision timescale.  
The parameter we use to
describe the collision rate is 
the number of collisions per planet orbit, $N_c$, for an individual particle; 
$N_c \approx P/t_{col}$, where $P$ is an planet orbital period.
The parameter $N_c$ is related to the normal disk optical depth by
$N_c \approx 18 \tau_n$ \citep{hanninen92}. 
Each planetary orbit the number of collisions is computed from
the total number of particles in the simulation and the collision rate.
The particles that receive collisions are chosen randomly from
those integrated.  These particles receive perturbations
to their momentum at a randomly chosen time during the orbit.

If a particle suffers a collision its
radial velocity is adjusted to damp the energy,
\begin{equation}
v_r \to \epsilon v_r
\end{equation}
where $-1 < \epsilon \le  1$. 
Here we have set $\epsilon =0$.
Had we not
restricted our simulations to zero eccentricity planets we
would have required the radial velocity perturbation
to depend on the mean set by the distribution of particles.
The tangential velocity is given a random kick
\begin{equation}
v_\phi \to v_\phi +  \Delta v
\end{equation}
where $\Delta v$ is chosen from a normal distribution 
with dispersion $\sigma_{dv}$.

Because the velocity perturbation at each collision,
$\Delta v$, is chosen from a normal distribution
its expectation is zero. 
While total
angular momentum should not drift rapidly,
angular momentum is not conserved during the simulation.
Adjustment or redistribution of angular momentum following collisions 
would be required to
conserve angular momentum (e.g., as explicitly described 
by \citealt{melita98}).
We do not chose to do this as we continually
regenerate particles
that have scattered outside the region of interest.
To conserve angular momentum
computational time would have been wasted integrating particles that diffuse
away from the planet as well as toward it.
Our adopted procedure allows us to
simulate a diffusive disk with minimal time spent
integrating particles outside the region near the
planet.

Particle initial semi-major axes were chosen to lie between
1.4 and 1.5 times the semi-major axis of the planet.
This way the particles were allowed to
diffuse from larger radius, approaching the planet.
Particles are removed from the simulation if they
have a semi-major axis greater than 1.6 
or less than 0.7, have an eccentricity greater than 0.5 or are 
unbound and if they move closer to the planet than 1/10 of its
Hill radius.   Particles that are removed from the simulation are 
regenerated choosing their orbital parameters in the
same way as the initial particles were chosen. This ensures
that a constant
number of particles is integrated at all times.  

The initial particle eccentricities were drawn from a Rayleigh distribution
with a mean eccentricity of $e_{init}$ that is chosen to
match the velocity perturbation parameter, $\sigma_{dv}$ or
\begin{equation}
e_{init} = { \sqrt{8}  \sigma_{dv} \over v_K}.
\label{eqn:e_init}
\end{equation}
where $v_K$ is the velocity of a particle undergoing circular motion
at the radius of the planet. 
A particle in a nearly circular orbit with 
eccentricity $e$ has radial velocity component 
${v_r\over v_K} = { e } \sin n t$ and
tangential component ${v_\phi \over v_K} = (1 + {e \over 2} \cos nt)$.
A distribution of particles with the same eccentricity
and random mean longitudes and longitudes of
periastron has radial dispersion
${<v_r^2>\over v_K^2} = {e^2 \over 2}$ 
and ${<(v_\phi-1)^2>\over v_K^2} = {<v_r^2> \over 4 v_K^2} = {e^2 \over 8}$.
This last expression accounts for the factor of $\sqrt{8}$ in equation
\ref{eqn:e_init}.  
We describe and refer to simulations
in terms of either the velocity perturbation, $\sigma_{dv}$, 
or the mean initial particle eccentricity, $e_{init}$, as they
are directly related to one another.

A steady state is reached after the disk has had time 
diffuse from the generation region (at a radius of 1.4-1.5) 
to the particle
removal region, near the planet's orbital radius.
We checked that the radial velocity dispersion distant
from the planet (greater than $r\sim 1.2$) is that expected
from perturbations 
given to the tangential velocity perturbation at each collision
or $<v_r^2> \approx 4 \sigma_{dv}^2$.
This is consistent 
with a distribution of particles with a mean eccentricity
of $<e> = e_{init} = {\sqrt{8} \sigma_{dv} \over v_K}$.
Because the radial velocity component is damped at each collision
the mean particle eccentricity does not
significantly increase past this value except 
near the planet.

A density distribution is created from particle positions
at the same time (planet mean anomaly) in each planetary orbit.
We have added 1000 of these from different timesteps
(each at the same time in the planet's orbit) to ensure
that the density distributions are smooth and have a large
dynamic range.
We work in units of the planet's semi-major axis, $a_p$, and orbital period.
The mass of the planet is described
in terms of its mass ratio, $\mu$, the ratio of the planet
mass to that of the central star.
We describe the velocity perturbation $\sigma_{dv}$ in units of the velocity of
a particle in a circular orbit, $v_K$, or in terms of the associated
initial particle eccentricity, $e_{init}$.

\section{Simulations}

The simulations we discuss have parameters listed in Table \ref{tab:tab1}.   
We relate the simulation parameters to 
quantities that can be constrained by observations, 
the disk normal optical depth, $\tau_n$, and the 
velocity dispersion that is related to the disk thickness.
We have run simulations for three different planet mass ratios,
$\mu = 10^{-4}, 10^{-5}$ and $10^{-6}$,  two different
collision rates,  $N_c = 10^{-2}$ and $10^{-3}$ collisions per
particle per orbit corresponding to disk normal optical depths 
$\tau_n =0.5\times 10^{-3}$ and $0.5\times 10^{-4}$ and four
different velocity perturbation sizes.  
In Table \ref{tab:tab1} the velocity perturbation sizes
are listed in terms of the initial mean eccentricity $e_{init}$ 
where $\sigma_{dv}/v_K =  e_{init}/\sqrt{8}$ and are related
to the disk velocity dispersion by $(u/v_K)^2 \approx e_{init}^2/2$. 

Azimuthally averaged density profiles (number density as a
function of radius)
after equilibrium is reached are shown in Figures 
\ref{fig:rplotp1} - \ref{fig:rplotc} where each
figure shows simulations with only one planet mass.
A comparison of Figure \ref{fig:rplotp1} - \ref{fig:rplotp3}
suggests that the radius at which the particle number density drops is
related to the planet mass.  

In the collisionless restricted 3 body system there
is an abrupt change in dynamics 
in the corotation region at a semi-major axis
that we refer to as the boundary of the chaotic zone.
The width of this
zone has been measured numerically and predicted theoretically
by predicting the semi-major
axis at which the first order mean motion resonances overlap
\citep{wisdom80,duncan89,murray97}.
In our figures we have with an arrow
marked the chaotic zone boundary at a radius $r_z = a_p + da_z$
with the distance between the planet and
chaotic zone $da_z = 1.5 a_p \mu^{2/7}$ 
\citep{wisdom80,duncan89} for the different simulations.  
We see from
the figures that this boundary is approximately the location at which the
radial profiles suffer a drop near the planet, as
would be expected from the change in the dynamics due to
overlap of mean motion resonances \citep{wisdom80,murray97,quillen06b}.

We note that the density drops before the chaotic zone
boundary for the simulations with the most massive planets.  
The simulations with the most massive planet and the steepest
profiles are also the ones that show the most structure. 
Figures \ref{fig:c2gray} and \ref{fig:f1gray} display
density distributions as a function of radius and angle
for simulations C2 and F1.   
We find for planet mass ratios equal to or below $10^{-5}$
little structure is evident in the density distribution.
However for a planet mass ratio of $10^{-4}$ and low
values of the velocity perturbation,
structure at strong mean motion resonances 
such as the 3:2 and 3:4 resonance is present.
When the resonances are strong trajectories 
can become planet orbit crossing.  Particles that
diffuse into these regions are removed from the simulation 
causing structure in the surface density (as seen in
figure \ref{fig:f1gray}) and a reduction in the density
profile at a radius larger than the chaotic zone boundary.
This is likely to account for the density drop past the chaotic zone boundary
evident in the profiles shown in figure \ref{fig:rplotp1}.


\renewcommand{\thefigure}{\arabic{figure}\alph{subfig}}

\setcounter{subfig}{1}
\begin{figure}
\includegraphics[angle=0,width=3.5in]{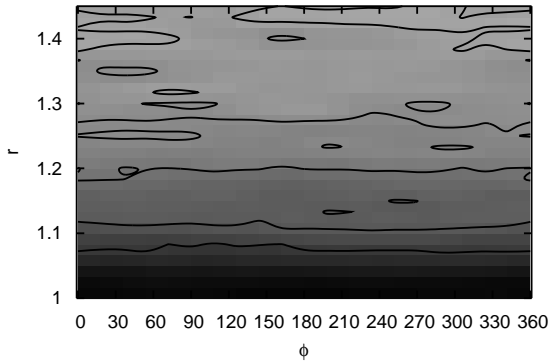}
\caption{
\label{fig:c2gray}
Number density as a function of $r$ and $\phi$ for
simulation C2 with planet mass ratio $\mu = 10^{-5}$.
Little structure is seen in the disk.
}
\end{figure}
\addtocounter{figure}{-1}
\addtocounter{subfig}{1}
\begin{figure}
\includegraphics[angle=0,width=3.5in]{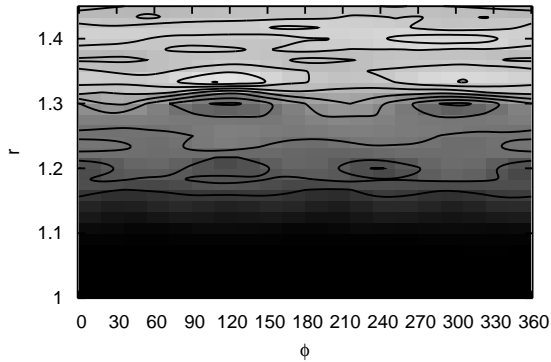}
\caption{
\label{fig:f1gray}
Number density as a function of $r$ and $\phi$ for
simulation F1 with planet mass ratio $\mu = 10^{-3}$.
Structure is seen in this disk associated with mean motion
resonances.   The 2:3 mean motion resonance
is located at $r=1.31$ and the 3:4 mean motion
resonances at 1.21. Both resonances are associated
with structure in the disk.
}
\end{figure}
\renewcommand{\thefigure}{\arabic{figure}}


In Figure \ref{fig:rplotr} we show profiles for the 3 simulations
with the same collision rate and perturbation
but different planet masses and with radius rescaled by
$\mu^{2/7}$ so that the $\mu=10^{-4}$ and $\mu=10^{-6}$ simulation
chaotic zone boundaries lie on that for the $\mu=10^{-5}$ simulation.
In other words $r$ has been rescaled in
this figure by a factor of $(10^{-5}/\mu)^{2/7}$.
Here we see that this radius does
give an estimate for the radius at which a disk
is likely to suffer a drop in density due
to removal of particles by the planet. 
The rescaled profiles do not lie on top of each
other; the lower mass planet simulations have shallower
profiles after rescaling.
The disk slopes are not self similar after rescaling implying
that the disk profile shape is not solely a function
of the distance to the chaotic zone.

The slope of the disk edge is dependent on
the collision rate and perturbation size as well as
the planet mass.  In figures 
\ref{fig:rplotc}, \ref{fig:rplotn}, and \ref{fig:rplotd},
we show profiles for the same number of collisions per
orbit and collision velocity perturbation
sizes but for different planet masses.
The simulations shown in \ref{fig:rplotn} have twice the
velocity perturbation size  and
the same collision rate as those shown in \ref{fig:rplotc}
whereas those shown in \ref{fig:rplotd} have the same
velocity perturbation size but a tenth the number
of collisions per orbit as those shown in \ref{fig:rplotc}.
We find that a larger change in the profile shape is
caused by variation in the velocity perturbation parameter
$\sigma_{dv}$ than variation by the same factor in the
collision rate $N_c$.

\begin{figure}
\includegraphics[angle=0,width=3.5in]{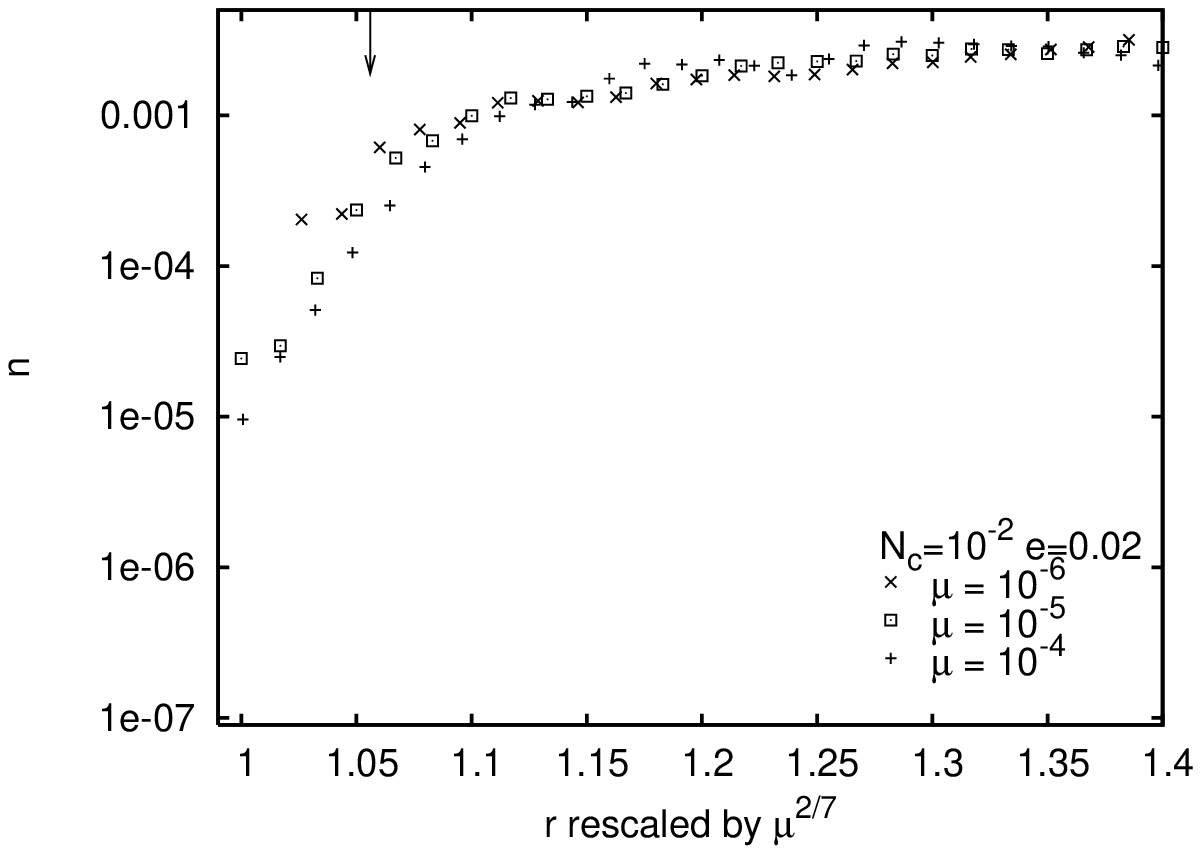}
\caption{
Azimuthally averaged density as a function
of radius for simulations with different planet
mass ratios C1,C2, and C3 but the same collision rate
and velocity perturbation size.
Here the radius has been rescaled by a factor of $(10^{-5}/\mu)^{2/7}$
so the chaotic zone boundaries for the three simulations line up.
\label{fig:rplotr}
}
\end{figure}

\subsection{Diffusive descriptions for the density drop}

We now discuss diffusive descriptions for
the disk edge to better understand the parameters
affecting the decrease in particle density near the planet.
Since the velocity dispersion is $u^2 \approx <v_r^2>$ and
$<v_r^2> \sim 4 \sigma_{dv}^2$  we estimate that the
diffusion coefficient (equation \ref{eqn:diffusion}) is 
\begin{equation}
D  \approx {4 \sigma_{dv}^2  \over t_{col} n^2} .
\end{equation}
In units of the square of the planet semi-major
axis divided by the rotation period,
\begin{equation}
D \approx 4 N_c \left({\sigma_{dv} \over v_K}\right)^2  
=  {e_{init}^2 N_c \over 2 }.
\end{equation}

The diffusion equation for an isolated
ring, Equation \ref{eqn:diffusion}, must be modified to
include the effect of the planet.  When
a steady state is reached (${\partial N \over \partial t}\approx 0$),
a simple model is 
\begin{equation}
{\partial  \over \partial r} \left(D {\partial N \over \partial r} \right)
\approx 
\left.{\partial N \over \partial t}\right|_{planet}  
,
\label{eqn:diffusionp}
\end{equation}
(e.g., \citealt{quillen06})
where the right hand side represents the rate
that particles are removed from radius $r$ 
because of perturbations from the planet. 
This term is expected to be a function of planet mass and
distance to the planet but could also be a function
of the particle velocity distribution or eccentricity distribution.

We assume that there is only one important scale in the problem,
the distance between the planet and the chaotic zone boundary,
$da_z = 1.5 \mu^{2/7}$.  We rewrite equation \ref{eqn:diffusionp}
with in terms of variable $y = {r-r_p \over da_z}$ and 
assuming that the diffusion coefficient is independent
of $r$ and $N$, 
\begin{equation}
{\partial^2  N \over \partial y^2} 
\approx 
{da_z^2 \over D} 
\left.{\partial N \over \partial t}\right|_{planet}  
.
\label{eqn:diffusionu}
\end{equation}
%
%
A simple form for the right hand side of equation \ref{eqn:diffusionp},
describing removal of particles by the planet, would be
\begin{equation}
\left.{\partial N \over \partial t}\right|_{planet}  \sim 
\left\{
\begin{array}{ll}
       {N f(y) \over t_{remove}} & {\rm for} ~~ 0 < y  <  1  \\ 
       0                         & {\rm for} ~~ y \ge  1
\end{array}
\right.
\label{eqn:case}
\end{equation}
When $f(y)=1$ the solution
\begin{equation}
N(y) \propto e^{yl}   
\label{eqn:nexp}
\end{equation}
with inverse scale length 
\begin{equation}
l \sim \sqrt{ da_z^2 \over t_{remove} D }.
\label{eqn:lsqrt}
\end{equation}
If $f(y) = 1-y$, then $N(y)$ is an Airy function  and if
$f(y) = \exp(-y)$ (as might
be expected from lifetime measurements in collisionless
integrations; \citealt{david03})
the solution is a modified Bessel function.
These solutions also behave in an exponential manner with an inverse
scale length that is of the same order of magnitude as $l$.
Figures \ref{fig:rplotp1} - \ref{fig:rplotp3}, 
\ref{fig:rplotc} - \ref{fig:rplotd} show
that the density profiles do approach 
an exponential form near the planet where the density gradient is steepest.
In parameters related to our simulations 
we expect the inverse scale length
\begin{equation}
l \sim   2
         t_{remove}^{-1/2} 
         \mu^{2/7} N_c^{-1/2} 
          e_{init}^{-1}.
\label{eqn:lexp}
\end{equation}

\renewcommand{\thefigure}{\arabic{figure}\alph{subfig}}
\setcounter{subfig}{1}
\begin{figure}
\includegraphics[angle=0,width=3.5in]{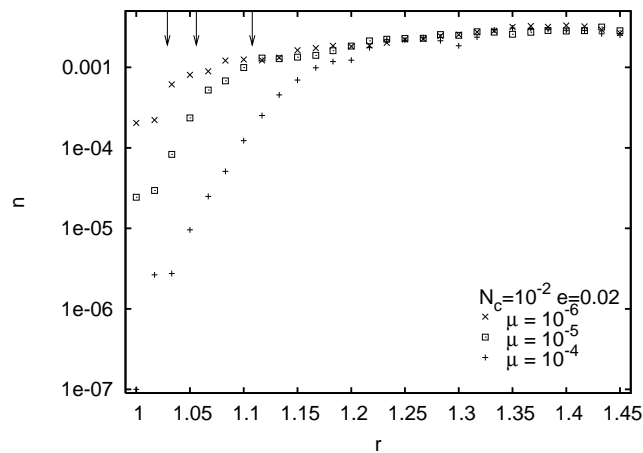}
\caption{
Azimuthally averaged density as a function
of radius for simulations with different planet
mass ratios 
but the same collision rate and velocity perturbation sizes.
Here simulations C1,C2, and C3  are shown.
\label{fig:rplotc}
}
\end{figure}

\addtocounter{figure}{-1}
\addtocounter{subfig}{1}
\begin{figure}
\includegraphics[angle=0,width=3.5in]{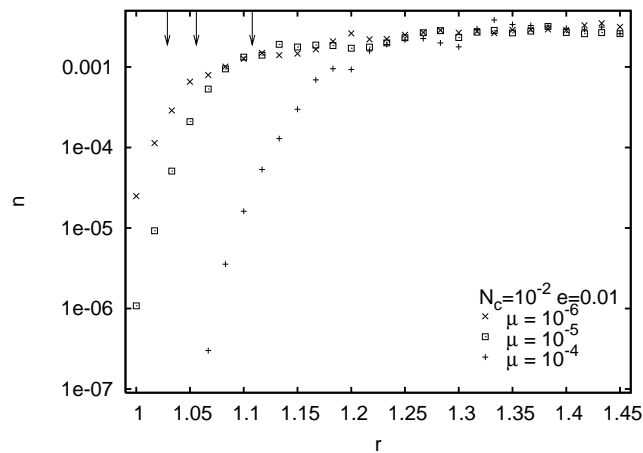}
\caption{
Same as figure \ref{fig:rplotc} but for simulations
N1,N2,N3.  
These simulations have twice the
velocity perturbation size  and
the same collision rate as those shown in \ref{fig:rplotc}.
\label{fig:rplotn}
}
\end{figure}

\addtocounter{figure}{-1}
\addtocounter{subfig}{1}
\begin{figure}
\includegraphics[angle=0,width=3.5in]{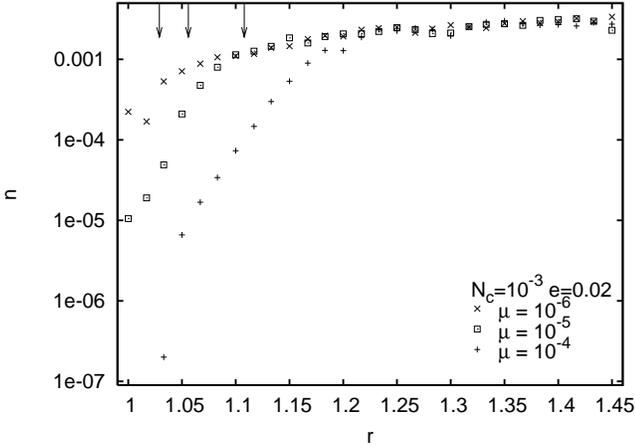}
\caption{
Same as figure \ref{fig:rplotc} but for simulations
D1,D2,D3. 
These simulations have the same 
velocity perturbation size but a tenth the number
of collisions per orbit as those shown in \ref{fig:rplotc}.
\label{fig:rplotd}
}
\end{figure}

\renewcommand{\thefigure}{\arabic{figure}}

From the simulations we measure the density
near the planet compared to that away from the disk edge
or specifically the log of this ratio
\begin{equation}
\Delta \equiv  \log_{10} \left({N(y=0) \over  N(y \ga 1)}\right).
\label{eqn:delta}
\end{equation}
Equations \ref{eqn:nexp} and \ref{eqn:lexp}
suggest
that $\Delta$ depends on parameters describing
our simulations, such as the planet mass
ratio, collision rate and velocity perturbation size,
however the above approximations and assumptions 
may be too simplistic to give an accurate prediction
for the inverse scale length $l$. Furthermore
$l$ contains the unknown timescale, $t_{remove}$.

The density decrement, $\Delta$,  measured from the radial profiles,
is plotted in Figures \ref{fig:lsimsn2} and \ref{fig:lsimsn3}.
As predicted by equation \ref{eqn:lexp}, 
our measured density decrement is consistent with a power
law function of the simulation parameters.  Also as we expected
$\Delta$ increases with increasing planet mass ratio,
and decreases with increasing 
collisions rate and velocity perturbation size. 
To measure the exponents, 
we have fit lines to the density decrements 
finding pretty good correspondence with the following function
that is also plotted in Figures \ref{fig:lsimsn2} and \ref{fig:lsimsn3},
\begin{eqnarray}
\log_{10} \Delta & \approx  &
   0.12 +
   0.23\log_{10} \left({\mu   \over 10^{-6}}\right)
  -0.1 \log_{10} \left({N_c  \over 10^{-2} }\right) \nonumber \\
&& ~~~ -0.45 \log_{10} \left({e_{init} \over 0.01 }\right)
\label{eqn:f1}
\end{eqnarray}
This function implies that the inverse scale length
\begin{equation}
l \sim 3.0   \left({\mu      \over 10^{-6}}\right)^{0.23} 
             \left({N_c      \over 10^{-2}}\right)^{-0.1}
             \left({e_{init} \over 0.01   }\right)^{-0.45}.
\label{eqn:lscale}
\end{equation}

A comparison between the measured scale length above 
(equation \ref{eqn:lscale})
and the predicted by equation \ref{eqn:lexp} reveals some inconsistencies.
Equation \ref{eqn:lexp} implies that the 
length scale should be proportional to $N_c^{-1/2} e_{init}^{-1} $
rather than $N_c^{-0.1} e_{init}^{0.45}$ as we find above.  
The above function measured $l$ combined 
with equation \ref{eqn:lexp} would imply that
\begin{equation}
t_{remove} \sim  145
                \left({\mu      \over 10^{-6}}\right)^{0.11} 
                \left({N_c      \over 10^{-2}}\right)^{-0.8} 
                \left({e_{init} \over 0.01}   \right)^{-1.1}
\label{eqn:trem}
\end{equation}
in orbital periods.
It is not clear why the removal timescale would increase with
planet mass. Had we used a larger scale for
the distance than the distance to the chaotic
zone  boundary in equations \ref{eqn:case} and \ref{eqn:lsqrt}
then the form of $t_{remove}$ would have changed. 
It is expected that $t_{remove}$ depends
on the velocity perturbation size as we do expect
the particle lifetime near the planet to depend
on particle eccentricity, with shorter lifetimes or removal
timescales at higher particles eccentricities.
Perhaps the particle lifetime
depends on the collision rate because particles
tend to diffuse to higher and higher eccentricity between collisions.
Our measured $l$ is not as strong a function of $N_c$
as predicted by equation \ref{eqn:lexp}.
A better theoretical model for both particle lifetime and
diffusion is required to better match the measured 
exponents of equation \ref{eqn:lscale}.

\subsection{Disk truncation}

The shallowest 
profiles displayed in Figures \ref{fig:rplotp1} - \ref{fig:rplotp2}
represent particle disks that have sufficiently high collision rates 
and velocity dispersions that a planet cannot efficiently
truncate the disk.
It is interesting to apply our measured density decrement
to estimate the minimum planet mass
that can truncate a disk.
We set a limit of $\log_{10} \Delta = 0.12$ (corresponding
to a density drop of 20) from the shallowest
simulations and solve equation \ref{eqn:f1}
to find the mass ratio of a planet that cannot effectively truncate
at disk of a particular collision rate and velocity dispersion.   
To truncate a low optical depth particle disk a planet
must have mass ratio greater than
\begin{equation}
\log_{10}\mu \ga -6
    + 0.43 \log_{10} \left({N_c \over 10^{-2}}\right)
    + 1.95 \log_{10} \left({e_{init} \over 0.01}\right)
\label{eqn:mulim}
\end{equation}
or in quantities more directly related to observations
\begin{equation}
\log_{10}\mu \ga -6 
    + 0.43 \log_{10} \left({\tau_n \over 5 \times 10^{-3}}\right)
    + 1.95 \log_{10} \left({u/v_K  \over 0.007}\right).
\label{eqn:mulimobs}
\end{equation}

The function (equation \ref{eqn:mulim}) is plotted
in figure \ref{fig:total} for collision rates 
$N_c=10^{-2}$ and $10^{-3}$ collisions
per orbit corresponding to disk optical depths 
$\tau_n =0.5\times 10^{-3}$ and $0.5\times 10^{-4}$.  
Instead of plotting this minimum mass ratio
as a function of $e_{init}$ we plot as a function of
velocity dispersion using $e_{init}^2 \sim 2 (u/v_K)^2$.
On this plot we also show a limit corresponding
to the eccentricity of a particle that would allow a particle
to cross from the chaos zone boundary to the planet's semi-major
axis, or the limit $(u/v_K) \sim 1.5 \mu^{2/7}$.  We refer
to this limit on the plot as $\mu_e$.

It is interesting to compare our mass limit for disk
truncation to that expected for
an $\alpha$ accretion disk.   A gaps opens
in a viscous accretion disk when the torque driven by 
spiral density waves excited by a planet exceeds viscous inflow or
\begin{equation}
\mu \ga 40 {Re}^{-1}
\end{equation}
\citep{lin79,bryden99,ward97}.
Here the inverse of the Reynold's number, $Re^{-1} = \alpha (u/v_K)^2$.
We have plotted this function on Figure \ref{fig:total} 
for $\alpha = 0.001$.
A particle disk has effective viscosity  set by the
diffusion coefficient of equation \ref{eqn:dcoeff}
(e.g., \citealt{M+D}), hence
if our particle disks behaved similar to an accretion
disk then this line (shown as a thin 
dotted line in figure \ref{fig:total})
would like approximately on top of the dotted line for $N_c = 10^{-2}$
with $\alpha \sim N_c/6$. 
It is interesting that
the disk truncation criterion scales with
velocity dispersion in the same way as the accretion disk; both require
that the planet mass ratio exceed a constant times
the square of the velocity dispersion.
We find that a larger planet is required to open a gap in an accretion 
disk with equivalent viscosity to 
the diffusion coefficient in a low optical depth particle disk.
This is somewhat counter-intuitive as spiral density waves
excited by a planet carry angular momentum pushing 
material away from the planet and they cannot
be driven in a low optical depth particle disk.
However collisions damp orbital eccentricity and high
eccentricity orbits can be more quickly scattered by a planet.
A lower mass planet might be able to truncate
a particle disk because the long timescale between
collisions allows particle eccentricities to grow near the planet.

Figure \ref{fig:total} illustrates that for particle
disks in the regime expected for debris disks, a Neptune
mass planet is sufficiently massive to truncate the
disk, and if the velocity dispersion is low 
($u/v_K \la 0.05$) and the disk optical depth low, $\tau_n \la 10^{-4}$,
an Earth mass planet suffices. 

\section{Discussion}

Here we have not considered models with dust produced
in resonance (e.g., \citealt{quillen02,wyatt03}) but
only allowed particles to diffuse inward toward resonances.
In resonance, certain angles with respect to the planet have
longer removal timescales than others. Effectively 
$\left. {\partial N \over \partial t} \right|_{planet}$
depends on the orbital elements of the particle distribution
and the diffusion equation should be expanded to depend
not on radius but all the orbital elements. 
Individual resonances may play an important role setting
the velocity dispersion disks near larger planets causing increased
diffusion near the disk edge (e.g., \citealt{quillen06,quillen06b}).
We expect additional complexity and structure if particles are produced
trapped in resonance or if the system is not in a steady state.
The simulations were run an average of $10^5$ orbits which
is long compared to the age of some large debris disks.
Our integration times are long because we allowed particles to diffuse
toward the planet from a moderately distant location and we
ran each simulation sufficiently long to achieve a steady state.
Additional phenomena may be discovered in evolving systems.

The simulations were carried out with
a collision rate and associated collision induced velocity perturbation
independent of position, however particles
near the planet should have suffered fewer collisions because of 
the steep drop in density.  Particles in the disk edge should have
also suffered larger velocity perturbations as the velocity
dispersion increases nearing the planet.  The velocity
dispersion increase (spanning a factor of a few
for the largest planet) is not as significant as the density  decrease
nearing the planet (spanning orders of magnitude for the largest planet
simulations).
A more accurate simulation probably would have a 
steeper disk edge due to the decrease in collision rate near the planet
and would have required longer to achieve equilibrium
as a consequence.

Here we have adopted a diffusive approach toward simulating
particle disks and neglected the role of the particle size distribution.
If collisions are destructive then 
diffusion caused by collisions can still take place. 
However each diffusive step involves a destructive collision
producing smaller bodies and these in turn have
a shorter collision timescale.    We imagine
an effective diffusion coefficient that depends not only on 
local particle density but on the size distribution
with the larger particles effectively having longer
collisions times and smaller diffusion coefficients.
The different sized particles also may
have different velocity dispersions.
In this case we would expect that the larger
bodies would have smaller diffusion coefficients
as the large bodies would have
lower velocity dispersions.
It is tempting though premature to predict that the different
size particles would exhibit different disk morphologies 
as a consequence of their different collisional timescales
and velocity dispersions.

It is interesting to compare our simulations to imaging studies
of debris disks.  Our lower mass limit (equation \ref{eqn:mulimobs}, see
circle in Figure \ref{fig:total}) is near but a few times lower than 
the previous limit set by \citet{quillen06}
for Fomalhaut.  As noted by \citet{quillen06b} the dynamics of low 
free eccentricity particles near an eccentric planet
is similar to low eccentricity particles near a planet in a
circular orbit, so we can extend equation \ref{eqn:mulimobs}
to low eccentricity planets.  Observations of Fomalhaut's disk
have not revealed any lumps or asymmetries in the disk, other than
its eccentricity \citep{kalas05}, consequently the planet mass
must not be high ($\mu \la 10^{-4}$). Otherwise Fomalhaut's disk would
look like the disk displayed in Figure \ref{fig:f1gray}.  Consequently
we also confirm, in a different way, the upper limit for
the planet mass proposed by \citet{quillen06}.
In contrast, Epsilon Eridani's disk does
show clumpy structure \citep{poulton06}.  
If particles are produced outside
resonance (as explored here) then we would infer that a moderate mass
planet and low dispersion disk might account for the morphology. 

\section{Summary}

We have simulated the effect of collisions on a particle disk  by 
introducing velocity perturbations into numerical integrations
of particles under the influence of gravity  from a star
and a single planet. 
We find that the planet's chaotic zone boundary is an approximate
location for the disk edge for disks that have optical depths
in the range of observed debris disks.
We find that density profiles do depend on the collision 
timescale but are more strongly dependent
on the velocity perturbation size or velocity dispersion in 
the disk.  The simulated disk morphology is axisymmetric and smooth
unless the planet mass ratio $\ga 10^{-5}$
and the velocity perturbation size $\la 0.02$.  For higher
mass ratio planets and lower velocity perturbation simulations
structure is seen in the vicinity of strong mean
motion resonances at locations where particles
are more quickly removed from the simulation due to interactions
with the planet.

From the simulations we have measured a density decrement
describing the log
of the density ratio of that close to the planet and that past
the disk edge.  The density decrement depends on powers 
of the planet mass ratio, collision timescale and velocity
perturbation size.  With a diffusion approximation this decrement
represents a scale length that might in future be predicted 
with better understanding of particle dynamics 
near the planet.

We have used our numerically measured density decrement
to predict the mass of a planet required to truncate
a particle disk (equation \ref{eqn:mulimobs}).  
The minimum planet mass required
depends on the square of the disk velocity dispersion,
similar to gap opening in an accretion disk.  However
we find that a lower mass planet can truncate
a low optical depth disk than an accretion disk with an equivalent
viscosity.    The minimum planet mass also depends
on the collision rate.   As the collision rate is
related to the disk optical depth and the velocity
dispersion related to the disk thickness, 
equation \ref{eqn:mulimobs}
can be used to estimate the mass of an unseen planet
from observations of debris disks.


\renewcommand{\thefigure}{\arabic{figure}\alph{subfig}}
\setcounter{subfig}{1}
\begin{figure}
\includegraphics[angle=0,width=3.5in]{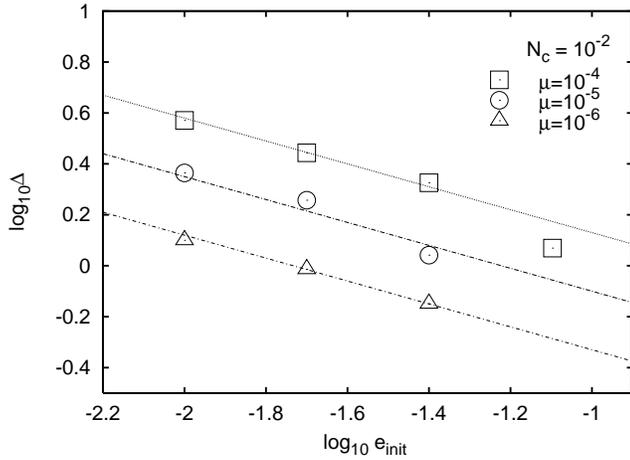}
\caption{
The log of the density decrement, $\Delta$, (related to the ratio of
the density past the chaotic zone to that near the planet, see equation 
\ref{eqn:delta}) 
versus the log of the velocity perturbation size.  
Only the simulations with $N_c = 10^{-2}$  collisions per orbit
are plotted.
Squares, circles and points represent simulations with
planet mass ratio $\mu=10^{-4}, 10^{-5}$ and $10^{-6}$ respectively.
Also plotted is a linear function described by equation \ref{eqn:f1}.
\label{fig:lsimsn2}
}
\end{figure}

\addtocounter{figure}{-1}
\addtocounter{subfig}{1}
\begin{figure}
\includegraphics[angle=0,width=3.5in]{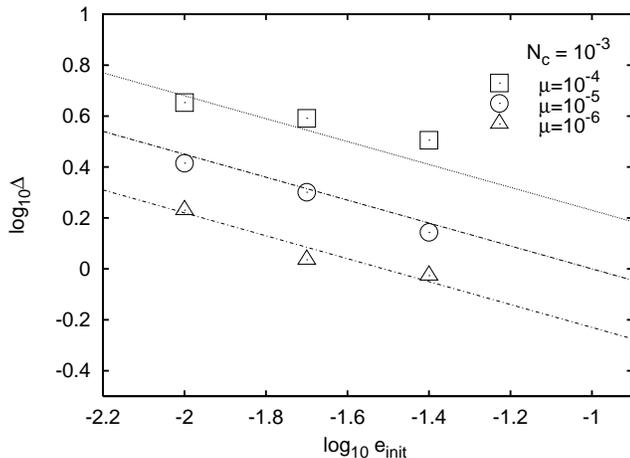}
\caption{
Same as Figure \ref{fig:lsimsn2} except for $N_c = 10^{-3}$.
\label{fig:lsimsn3}
}
\end{figure}

\renewcommand{\thefigure}{\arabic{figure}}

\vskip 0.1truein
Support for this work was in part
provided by National Science Foundation grant AST-0406823,
and the National Aeronautics and Space Administration
under Grant No.~NNG04GM12G issued through
the Origins of Solar Systems Program,
and HST-AR-10972 to the Space Telescope Science Institute.

\begin{figure}
\includegraphics[angle=0,width=3.5in]{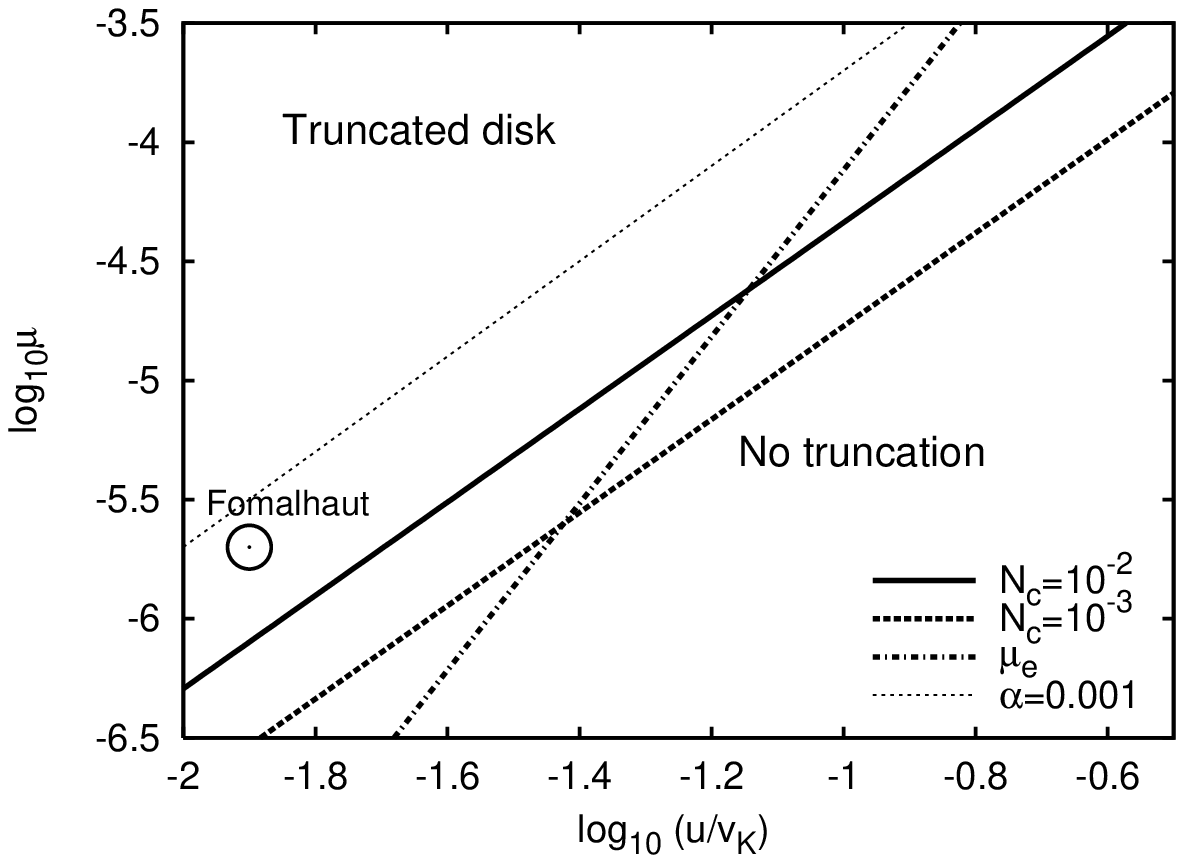}
\caption{
Disks can be truncated by planets with a mass ratio
that lies above the lines drawn.  
The $x$-axis is the disk velocity dispersion.
The solid and dashed line refer to particle disks with a limit set
by equation \ref{eqn:f1} for $N_c = 10^{-2}$ and $10^{-3}$
collisions per particle per orbit, respectively.
The thin dotted line is the gap opening criteria expected
for an $\alpha$ accretion disk with $\alpha = 0.001$.
The dot dashed line is a limit set by requiring that the
disk thickness be lower than chaotic zone width. 
A circle has been placed at the estimated planet mass lower limit for
a planet inside
Fomalhaut's disk edge using $N_c  = 3 \times 10^{-2}$ 
expected from the normal optical depth,
$\tau_n = 1.6 \times 10^{-3}$ \citep{marsh05} 
and velocity dispersion estimated from
the disk edge slope $u/v_K \sim 0.013$ \citep{quillen06}.
\label{fig:total}
}
\end{figure}

\vfill\eject
\clearpage

\begin{table*}
\begin{minipage}{120mm}
\caption{List of Simulations}
\label{tab:tab1}
\begin{tabular}{@{}lccccc}
\hline
Name &  $\mu$     & $N_c$     & $e_{init}$ \\
C1   &  $10^{-4}$ & $10^{-2}$ & 0.02  \\ 
C2   &  $10^{-5}$ & $10^{-2}$ & 0.02  \\
C3   &  $10^{-6}$ & $10^{-2}$ & 0.02  \\
D1   &  $10^{-4}$ & $10^{-3}$ & 0.02  \\ 
D2   &  $10^{-5}$ & $10^{-3}$ & 0.02  \\ 
D3   &  $10^{-6}$ & $10^{-3}$ & 0.02  \\ 
E1   &  $10^{-4}$ & $10^{-3}$ & 0.04  \\ 
E2   &  $10^{-5}$ & $10^{-3}$ & 0.04  \\ 
E3   &  $10^{-6}$ & $10^{-3}$ & 0.04  \\ 
F1   &  $10^{-4}$ & $10^{-3}$ & 0.01  \\ 
F2   &  $10^{-5}$ & $10^{-3}$ & 0.01  \\ 
F3   &  $10^{-6}$ & $10^{-3}$ & 0.01  \\ 
M1   &  $10^{-4}$ & $10^{-2}$ & 0.04  \\ 
M2   &  $10^{-5}$ & $10^{-2}$ & 0.04  \\ 
M3   &  $10^{-6}$ & $10^{-2}$ & 0.04  \\ 
N1   &  $10^{-4}$ & $10^{-2}$ & 0.01  \\ 
N2   &  $10^{-5}$ & $10^{-2}$ & 0.01  \\ 
N3   &  $10^{-6}$ & $10^{-2}$ & 0.01  \\ 
L1   &  $10^{-4}$ & $10^{-2}$ & 0.08  \\ 
\hline
\end{tabular}
{ \\
The parameters describing the discussed simulations are as follows:
The planet mass ratio is $\mu$, whereas
$N_c$ is the average number of collisions suffered
by each particle per planet orbit.
The dispersion of the azimuthal velocity kick given each collision,
$\sigma_{dv}/v_K =  e_{init}/\sqrt{8}$  
in units of the velocity of a particle
in a circular orbit.
The planet was in a circular orbit.
During each collision event the radial 
velocity component is set to zero
($\epsilon=0$).
Particles were removed from the simulation
if semi-major axis $a>1.5$ or $a<0.7$, eccentricity
$e>0.5$ or the particle passed within 0.1 Hill radii of
the planet.  Particles were generated and regenerated
when removed with a semi-major between 1.4 and 1.5 and
a mean eccentricity of $e_{init}$.
Simulations were run $10^5$ planetary orbits excepting
the F series that were run 
twice as long and the L1 simulation that was run half as long.
These times are long because we allowed particles to diffuse
toward the planet from a moderately distant location and we
ran each simulation sufficiently long to achieve a steady state.
}
\end{minipage}
\end{table*}

\end{document}